\documentclass[preprint]{revtex4-2}   

\usepackage{soul}
\usepackage{xr-hyper}   
\usepackage{hyperref}
\usepackage{amsmath,amssymb}
\usepackage{bbold}
\usepackage{bm}
\usepackage{graphicx} 
\usepackage{grffile}
\usepackage{microtype}
\usepackage{placeins}

\usepackage[parfill]{parskip}   

\usepackage{physics}

\usepackage{xcolor}   

\usepackage{mathtools}     
\usepackage[makeroom]{cancel}  

\usepackage{empheq}   
\usepackage{comment}  

\newcommand{\eq}[1] {\begin{equation}{#1}\end{equation}}
\newcommand{\eqlabel}[2] {\begin{equation}\label{#1}{#2}\end{equation}}
\newcommand{\aleq}[1] {\begin{align}{#1}\end{align}}

\newcommand{\transp}[1]  {\qty({#1})^T}

\newcommand{\nhat} {\hat{\vb n}}
\newcommand{\that} {\hat{\vb t}}

\newcommand{\bsigma} {\vb*{\sigma}} 

\newcommand{\smallat}[1] {\big|_{#1}}

\renewcommand{\grad} {\nabla}
\renewcommand{\div} {\nabla\vdot}

\begin{document}

  \title{Collective multicellular patterns arising from cadherin-linked cytoskeletal domains}
  \author{XinXin Du}
  \affiliation{Center for Computational Biology, Flatiron Institute, New York, New York 10010, USA}
  \author{Ido Lavi}
  \affiliation{Center for Computational Biology, Flatiron Institute, New York, New York 10010, USA}
  \author{Michael J. Shelley}
  \affiliation{Center for Computational Biology, Flatiron Institute, New York, New York 10010, USA}
  \affiliation{Courant Institute of Mathematical Sciences, New York University, New York, New York 10012, USA}
  \maketitle

\section{Abstract}
In multicellular systems, adhesion complexes, such as those composed of E-cadherin and associated catenins, mechanically couple neighboring cells by directly linking their actin-based cytoskeletal assemblies. However, the mechanics of how forces are transmitted across these adhesions remains largely unstudied. Here, we introduce a biophysical model that explicitly couples adhesion complex dynamics to intracellular mechanics across cell boundaries.  A cadherin dimer plus associated catenins connecting two cells is represented as a spring whose ends experience drag with respect to the moving actin cytoskeleton.  The cytoskeleton is modeled as a contractile gel driven by myosin activity in its bulk and forces from adhesion on its boundaries. Our model captures this bidirectional coupling via a coarse-grained continuum framework and reveals a range of observed cell- and tissue-scale behaviors. These include global cell polarization of the multicellular collective, other polarization patterns and oscillatory dynamics, spontaneously formed actin rings within cells, and supracellular stress chains. Many of these features arise from modeling the direct mechanical coupling between cytoskeleton and adhesion.  This model can be extended to other adhesion–cytoskeleton feedback systems and used to advance our understanding of multicellular tissue dynamics, particularly during development.

\section{Introduction}
Cells physically interact with each other and with their environment through dimerized transmembrane adhesion molecules and their cytoskeletal partners. Far from strictly providing a simple attraction that keeps cells confluent in tissues, these adhesion complexes have their own internal dynamics and are also coupled directly into the active intracellular structures.

Various collective multicellular patterns observed in tissues suggest a high degree of coordination between the actin-based cytoskeletal assemblies of neighboring cells. For example, during follicle cell rotation in \emph{Drosophila}, the actin populations of individual cells polarize in the same direction prior to the onset of rotation \cite{Popkova2021,Viktorinova2009,Viktorinova2017,Barlan2017}. Stable cortical actin rings have also been observed in multicellular collectives \cite{Lecuit2007,Kasza2011,Koppen2006,Choi2013,Cavey2008,Padash2005}. In tissues such as the \emph{Drosophila} mesoderm \cite{Yevick2019} and the mouse intestine \cite{Barai2025}, researchers have identified supracellular actomyosin structures characterized by ``star-like'' patterns that span multiple cells. Other collective states include alternating polarized actin populations in \emph{Scaptodrosophila} \cite{Osterfield2023} and transient stress chains \cite{Lopez2020} in the \emph{Drosophila} thorax.  While most attention has focused on juxtacrine signaling pathways as the primary drivers of such multicellular coordination, mechanical coupling via adhesion complexes like those involving E-cadherin are also thought to play a role. Indeed, cadherins and associated catenins have been shown to be important for collective behaviors such as durotaxis \cite{Sunyer2016} and the precise organization of multicellular structures, such as ommatidia \cite{Founounou2021,Kafer2007}. However, it remains unknown whether the physical linking provided by adhesion complexes can directly affect multicellular patterns of actomyosin.   

In this paper, we model 
the physical linking provided by adhesions and investigate the emergence of global actomyosin patterns like those observed experimentally.  Previous theoretical works on cadherins have addressed phenomena such as cadherin clustering, differential adhesion, and force-rate binding relationships \cite{Quang2013,Thompson2020,Yu2022,Kafer2007,Buckley2014,Sens2020}.  While many multicellular models approach cell-cell coupling with trapping potentials or steric interactions \cite{Du2017,Drasdo1995,Sepulveda2013,Wenzel2021}, we emphasize the well-known fact, often ignored by models, that adhesion complexes, such as ones associated with E-cadherin, bind directly to the actomyosin assemblies in cells. Our model of this interaction is cast in a simple scenario where cells are fixed two-dimensional domains.  We specify the nature of microscopic interactions between the actomyosin population in the cell bulk and the adhesion population on the cell boundary, and we use principles of conservation to pose the resulting equations of motion.

Through analysis and finite element simulation, we show that our model can predict surprisingly complex dynamical behaviors and generate collective supracellular structures reminiscent of biological observations. In certain parameter regimes, the model produces stable global polarization and anti-polarization. In other regimes, we find stable cortical actin rings that become more prevalent as adhesion levels increase. In addition, the model exhibits transient actin stress cables like those observed in the \emph{Drosophila} thorax, along with more exotic behaviors such as oscillatory and chaotic actomyosin dynamics. Overall, our study provides a theoretical framework for understanding collective actomyosin behavior in multicellular systems with adhesion-based coupling.

\section{Model}
In modeling actomyosin assemblies interacting via cadherin-based adhesions, we describe the coupled dynamics of contractile gels in confluent two-dimensional (2D) cell domains and adhesion populations along their shared one-dimensional (1D) boundaries. Actomyosin gels in adjacent cells generate active contractile forces that can move and stretch adhesion complexes. In turn, as these adhesion complexes are moved and stretched, they exert drag forces on these gels and transmit mechanical forces between them by storing and releasing elastic energy.

Each cell contains two actin populations, one, a background population of passively crosslinked filaments having density $\rho_0$, and the other, a population of filaments both passively and actively crosslinked having density $\rho$. Assuming that these populations are in interchange with one another, with nucleation and degradation rates of the background actin fast compared to the rates of interchange, then $\rho_0$ can roughly be assumed to be constant; see section (\ref{S-sec:rho0}) in the SM. This background pool of actin has indeed been shown to exist in some systems \cite{Dehapiot2020}. We impose mass conservation and momentum balance for the active component, and specify a constitutive relation defining its material stress:
\begin{align}
    & \pdv{\rho}{t} + \div(\vb u\rho) = D\laplacian \rho + \nu-b\rho \quad, \label{eq:rho}\\
    & \vb 0 = \div \bsigma -\gamma(\rho+\rho_0)\mathbf u \quad, \label{eq:u} \\
    & \bsigma = \eta(\rho+\rho_0)^2\qty(\grad \vb u + \transp{\grad \vb u}) - s_1(\rho+\rho_0)^2\mathbb 1 +s\rho\mathbb 1 \quad .\label{eq:stress}
\end{align}
Here, $\vb u$ is the velocity of both populations of actin, $D$ is a diffusion coefficient, and $\nu$ and $b$ are conversion rates to and from the background actin population (with $\nu\propto\rho_0$). The force $-\gamma(\rho+\rho_0)\vb u$, with $\gamma$ a constant, represents drag on actin motion against an unmodeled substrate.  The quantity $\eta(\rho+\rho_0)^2$ is the viscosity of the actin gel, with the quadratic dependence on density discussed in \cite{Furthauer2021}.  The isotropic tensor $-s_1(\rho+\rho_0)^2\mathbb 1$ represents a steric repulsion between actin polymers, and $s\rho\mathbb 1$ models an isotropic contractile stress induced solely by the active material. The latter is the active term in the model.  Similar forms to Eqns.~\eqref{eq:rho}-\eqref{eq:stress} have been explored in various descriptions of actomyosin \cite{Foster2015,Hannezo2015,Singha2025,Furthauer2021}.   

We impose no-penetration on $\vb u$ and a shear-stress boundary condition:
\aleq{
\vb u\cdot \nhat\smallat{\Gamma} = 0 \quad ; \quad \that\cdot (\bsigma\,  \nhat)\smallat{\Gamma} = f[\psi] \quad, \label{eq:gelbc}
} 
where $f[\psi]$ indicates a 
force density in the tangent direction 
that is functionally dependent on the number density of linkers $\psi$ on the cell boundary $\Gamma$, which we describe below. We also impose zero diffusive flux on $\rho$:
\eqlabel{eq:rhobc}{
-D\grad \rho \cdot \nhat \smallat{\Gamma} = 0 \quad .
} 
Here $\nhat$ is the unit outward pointing normal and $\that$ is the unit tangent vector to $\Gamma$, required to be in the same direction for both cells bordering $\Gamma$.

\begin{figure}[ht]
   \centering
   \includegraphics[width=1.0\textwidth]{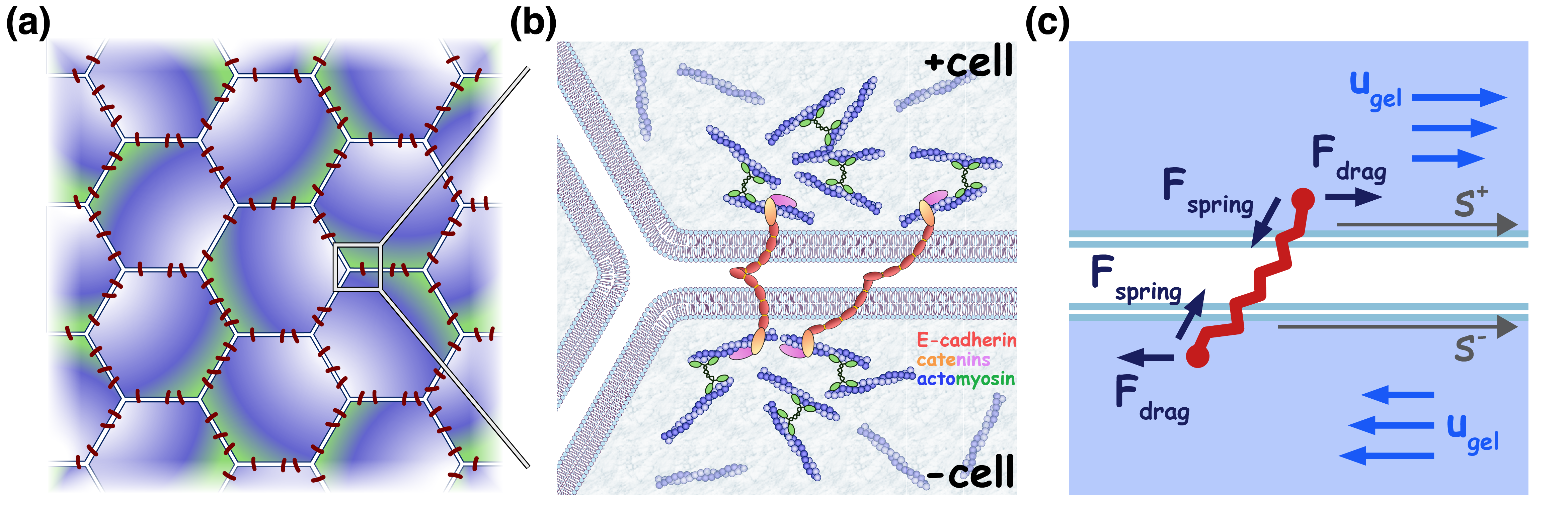} 
   \caption{(a) Schematic of confluent epithelium with hexagonal cells and linkers across cell membranes. (b) Schematic of E-cadherin transmembrane molecules (red) bound to each other and to catenins (orange and pink), which can in turn bind to cytoplasmic actin (blue), itself contracted by myosin (green). (c) An adhesion complex made of dimerized E-cadherin and associated catenins is modeled as a spring (red) whose ends experience drag forces with respect to gel velocities in neighboring cells (blue vector fields) and a restoring spring force. Arc length coordinates along the edges of ``+cell'' and `-cell'' are denoted $s^+$ and $s^-$, respectively.}
   \label{fig:intro}
\end{figure}

We describe, on each cell boundary, a population of double-bound linkers that mechanically couple the gels in adjacent cells. Here, a \emph{linker} represents the complex of dimerized cadherin plus its associated catenins.  Denoting the two cells as “$+$” and “$-$”, we let $s^+$ and $s^-$ represent the arc length labels of the linker ends on the sides of each respective cell.  Then, we define $\psi(s^+,s^-,t)$ as the number density of linkers attached between coordinate $s^+$ on the boundary of the ``$+$cell'' and coordinate $s^-$ on the boundary of the adjacent ``$-$cell'' at time $t$.  The evolution of $\psi$ is given by the Fokker-Plank equation:
\eqlabel{eq:FP}{
\pdv{\psi}{t} + \pdv {s^+}(v^+\psi) + \pdv {s^-}(v^-\psi) = D_\text{e}\qty(\frac{\partial^2 \psi}{\partial s^{+2}} + \frac{\partial^2 \psi}{\partial s^{-2}}) + k_\text{on}(|s^+-s^-|) -k_\text{off}(|s^+-s^-|) \psi \quad .
}
Here $v^+$ and $v^-$ are the velocities, along the cell sides, of the linker ends located at $s^+$ and $s^-$, respectively, and $D_\text{e}$ is their (1D) diffusion coefficient.  The nucleation and disassociation rates of doubly-bound linkers are modeled by the functions $k_\text{on}$ and $k_\text{off}$.  We assume that $k_\text{on}(|s^+-s^-|)$ is sharply peaked near zero (within a small typical length $\ell_\text{N}$), reflecting that linkers nucleate in an unextended configuration. We further forbid nucleation beyond a maximal extension $\ell_M$ (always greater than $\ell_N$). The unbinding rate $k_\text{off}$ is taken to be constant up to 
$\ell_\text{M}$, beyond which it increases rapidly to model the quick dissociation of overstretched linkers.  In particular,
\begin{align}
\left\{
\begin{aligned}
k_\text{on} &= a\left(e^{-\frac{|s^+ - s^-|}{\ell_\text{N}}} - e^{-\frac{\ell_\text{M}}{\ell_\text{N}}}\right)\,, \quad & k_\text{off} &= \lambda_0\,,
& \text{for } |s^+ - s^-| \leq \ell_\text{M} \\
k_\text{on} &= 0\,, \quad & k_\text{off} &= 10^5 \lambda_0\,,
& \text{for } |s^+ - s^-| > \ell_\text{M}  \label{eq:onoff}
\end{aligned}
\right.
\end{align}
where $a$ and $\lambda_0$ are constants. 

Expressions for $v^\pm$ are determined by microscopic force balance on linker ends. Each linker end is embedded in the local actin network and connected to its paired end by a linear spring.  Thus, it experiences both drag from the gel and a restoring spring force: 
\eqlabel{eq:eom}{
0=-\mu_0 (\rho^\pm(s^\pm)+\rho_0)\qty(v^\pm-u^\pm(s^\pm)) + k(s^\mp-s^\pm)  \quad ,
}
where $\rho^\pm(s^\pm)+\rho_0$ and $u^\pm(s^\pm)$ indicate the total gel density and tangent velocity in the {$\pm$cell} evaluated on the boundary.  
Through Eq.~\eqref{eq:eom} we obtain:
\eqlabel{eq:v}{
v^\pm(s^+,s^-) = \frac{k(s^\mp-s^\pm) }{\mu_0 (\rho^\pm(s^\pm)+\rho_0)} + u^\pm(s^\pm) \quad .
}
Equation~\ref{eq:eom} also defines the equal and opposite forces that linkers exert on the actin. The drag force applied on actin by a single linker (with ends at $s^+$ and $s^-$) at the point $s^+$ is ${\mu_0(\rho^+
+\rho_0)(v^+
-u^+
)=k(s^--s^+)}$. Similarly, the drag force applied at the point $s^-$ is $k(s^+-s^-)$. To obtain the total drag force at $s^+$ or $s^-$, we integrate the force density over positions on the other membrane ($s^-$ or $s^+$). That is,
\eqlabel{eq:fongel}{
f^+[\psi](s^+) = \int ds^- \, k(s^--s^+)\psi(s^+,s^-)  \quad , \quad 
f^-[\psi](s^-) = \int ds^+ \, k(s^+-s^-)\psi(s^+,s^-)\quad , 
} 
which sets the boundary condition in Eq.~\eqref{eq:gelbc}.

Lastly, we assume no-flux boundary conditions on Eq.~\eqref{eq:FP} at the cell tri-junctions ${s^\pm=0}$ and ${s^\pm=L_s}$; see Fig.~\ref{fig:intro}(a): 
\eqlabel{eq:linkerbc}{
-D_\text{e}\pdv{\psi}{s^+} + v^+\psi\smallat{s^+=0,L_s} =  -D_\text{e}\pdv{\psi}{s^-} + v^-\psi\smallat{s^-=0,L_s} = 0 \quad .
}
Notably, the evolution of the linker number density $\psi(s^+, s^-, t)$ depends on the state of the active gel in the cell interiors through the local actin densities $\rho^\pm(s^\pm)$ and velocities $u^\pm(s^\pm)$. Conversely, the dynamics of the active gel within each cell are influenced by the linkers via the boundary condition in Eq.~\eqref{eq:gelbc}, with $f[\psi]$ specified in Eq.~\eqref{eq:fongel}. Together, this coupled system enables mechanical communication between cells, mediated by the dynamic population of linkers transmitting forces across their boundaries.

To render our equations adimensional, we rescale time by the lifetime $\tau=1/b$ of polymerized actin, length by $\ell_0= \sqrt{D/b}$, and force by $F_0= \eta\, b^{5/2}/D^{3/2}$ (this form arises from the viscosity's quadratic dependence on density). We further scale the gel density $\rho$ in units of $\rho_0$, and the number density of linkers $\psi$ in units of $1/(\ell_0^2k')$, where $k'=k \ell_0/F_0$. The adimensional equations are then effectively given by setting $D,b,\eta,\rho_0$, and $k$ to 1, with the remaining parameters being adimensional control parameters; see section (\ref{S-sec:nondim}) of the SM.

\section{Results}
\begin{figure}[tbp]
   \centering
\includegraphics[width=0.9\textwidth]{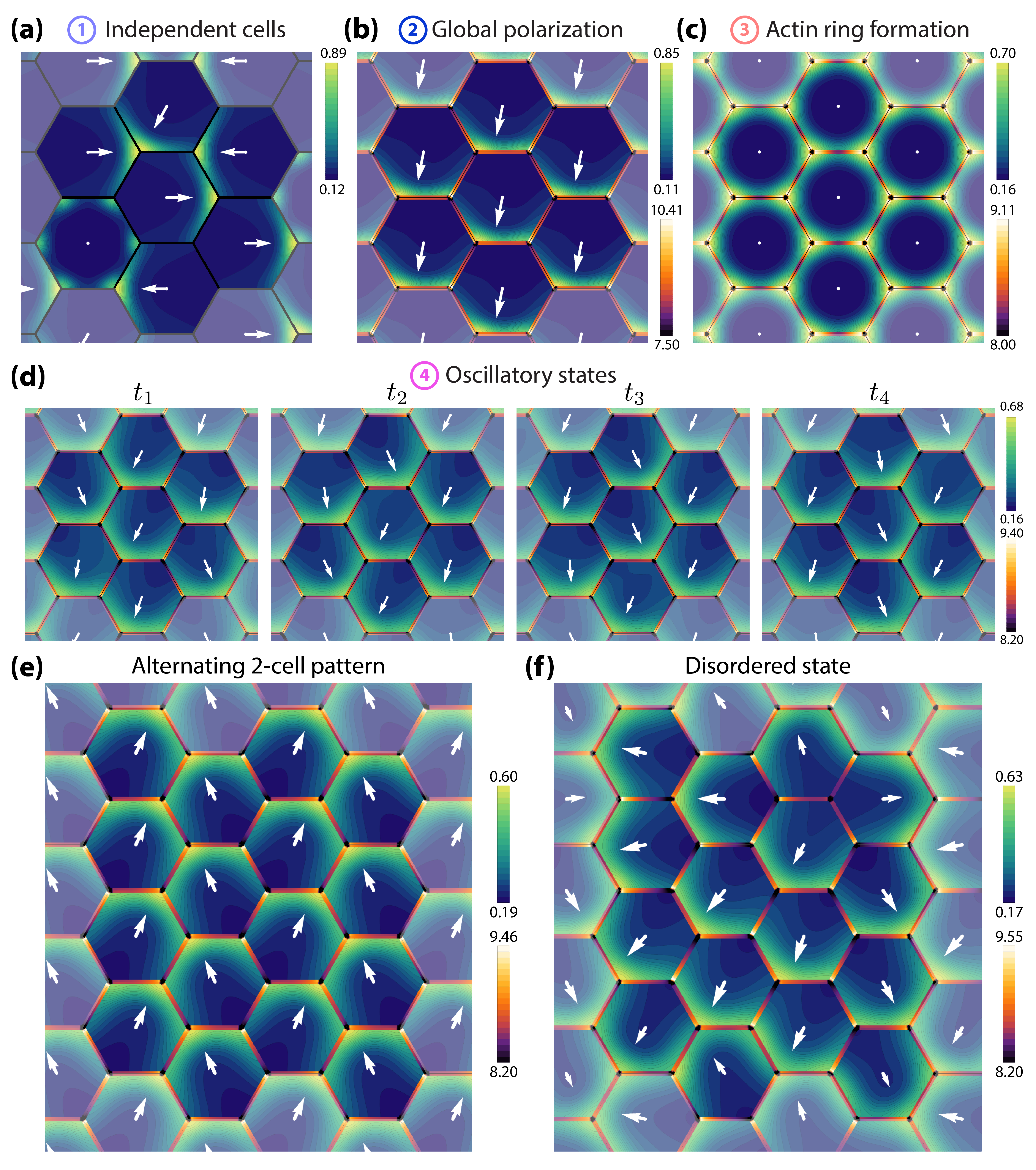} 
   \caption{The varieties of system behavior. (a) Steady state from simulation of actomyosin populations in isolated cells, i.e. no linkers between them. (b-d) Simulations of actomyosin populations coupled by adhesion, for a periodic tiling of the plane using the heptahex of cells. Panels (b-c) show globally coordinated, steady states: (b) global polarity, and (c) actin rings. Panel (d) shows a time course of persistent oscillations. (e) A steady state of alternating polarity in a 16-cell polyhex tiling of the plane.  (f) An example of a ``disordered'' steady state in the 16-cell system. Encircled numbers 1-4 correspond to labels in Figure 3. Gel parameters (a-f) $s_1=30$, $\gamma=0.001$, $L_s=8$; (a-b) $s=100$, $\nu=0.2$; (c) $s=100$, $\nu=0.3$; (d-f) $s=95$, $\nu=0.3$. Linker parameters (b-f) $a=80$, $\lambda_0=10$, $D_\text{e}=0.1$, $\mu_0=1$, $\ell_\text{M}=0.6$, $\ell_\text{N}=0.1$. }
   \label{fig:fig2}
\end{figure}

Cells in confluent tissues often exhibit hexagonal packing, with each cell having, on average, six neighbors. We mimic this geometry by simulating our model on a hexagonal array of cells that tiles the $xy$ plane. The minimal contiguous repeating tiling of regular hexagons in which each hexagon has six distinct neighbors is a heptahex; see Fig.~\ref{fig:fig2}(a-d) and section \ref{S-sec:tilings} of the SM. This tile consists of 7 regular hexagons joined along 21 shared edges.  In each hexagonal cell domain, we evolve an active gel problem using Eqns.~\eqref{eq:rho}--\eqref{eq:gelbc}. Note that the boundary condition Eq.~\eqref{eq:gelbc} is given by the linker number density $\psi$ using Eq.~\eqref{eq:fongel}. On each shared edge, we solve the Fokker-Planck problem given by Eqns.~\eqref{eq:FP} and \eqref{eq:linkerbc}. The velocities of linker ends, required for evolving $\psi$, are computed from the gels' densities and velocities via Eq.~\eqref{eq:v}. We numerically solve our coupled equations using a Finite Element Method (FEM), implemented with the open-source software FreeFEM++ \cite{Hecht}, with further details provided in section \ref{S-sec:computational} of the SM.

\FloatBarrier

We generally start our simulations from a nearly flat gel density in each cell, given by the balance of nucleation and degradation, plus a small perturbation such that $\rho=\nu+\delta\rho$ (with $\nu$ the dimensionless nucleation rate). We choose $\delta\rho$ as the sum of 20 randomly placed Gaussians with random small amplitudes and variances. We call these ``random initial configurations''. The linker number density $\psi$ is initially set to zero on all edges. 
Our rationale for selecting parameter values is discussed in section \ref{S-sec:params} of the SM.

In Fig.~\ref{fig:fig2}, we show representative simulations elucidating different system behaviors at long times. We plot gel densities in cell domains and linker end densities along cell edges.  Linker end densities are defined by $n^\pm(s^\pm)=\int_0^{L_s}ds^\mp\psi(s^+,s^-)$ which  computes the linker density $n^\pm$ at one linker end $s^\pm$ by summing over all possible positions of the other linked end at $s^\mp$.  White arrows in each cell indicates the polar order $\vb p$ of that cell, defined in Eq.~\eqref{eq:pH} below.  Figure~\ref{fig:fig2}(a) and Movie S1 shows the typical behavior of isolated cells without linkers (meaning $\psi\equiv0$ and accordingly $f[\psi]=0$ in Eq.~(\ref{eq:gelbc})). Gels in these cells form condensates that typically undergo spontaneous symmetry breaking, each independently localizing to the edges or corners of their domain. Additionally, we found infrequent instances where a gel in an isolated cell stabilized into a 6-fold symmetric state (lower left cell in Fig.~\ref{fig:fig2}(a) and Movie S1). These behaviors are driven strictly by the internal contractility of the active gel in the bulk of each cell. The formation of condensates above a critical contractility strength $s$ in the gel is consistent with our linear stability analysis in a simplified geometry (Appendix) and with other reports of edge-localized patterns in related systems \cite{Rautu2025}. 

Simulations of the fully coupled system reveal globally coordinated steady states that are reminiscent of biological observations. Figure~\ref{fig:fig2}(b) and Movie S2 show a steady state of a system with the same gel parameters and similar random initial conditions as in Fig~\ref{fig:fig2}(a), but with dynamic adhesions now in play. In this state, all cells polarize in the same direction due to linker-mediated force transmission across boundaries. This type of collective behavior is reminiscent of observations in follicle cells \cite{Popkova2021}.  Another global state is shown in Fig.~\ref{fig:fig2}(c) and Movie S4, where the nucleation rate, $\nu$, of the gels is higher than that in (b). Here, all gels stabilize to a 6-fold symmetric state, reminiscent of actin ring shapes in cells with stable actin cortices. Though isolated cells do occasionally and independently form ring states (Movie S3), our simulations suggest that linkers promote the coordinated spread of this pattern across the cell collective.  The influence of linkers on the spreading of the actomyosin ring pattern across multiple cells is most apparent in the simulation in Movie S5, where a larger system of 16-cells (see below) and a lower contractility strength $s$ are are considered.

Steady state patterns are not the only possible attractors in this system. There are parameter regimes in which oscillatory states emerge spontaneously. For example, the simulation in Fig.~\ref{fig:fig2}(d) and Movie S6 have the same parameters as in (c) but with lowered contractility strength $s$. This simulation shows gel polarity shifting sequentially across adjacent cells, and suggests the presence of a traveling wave propagating across the tile; the traveling wave is most evident in Movie S6, where the 7-cell repeating unit is visualized in a zoomed-out view. Actomyosin contractility waves are frequently reported in developing tissues \cite{Bailles2019,Bailles2022}, raising the possibility that our simulated polarity wave may reflect a similar mechanism.

We next explored an alternative periodic tiling of the plane that increased the system size. We selected a contiguous tile composed of 16 hexagons joined along 48 shared edges. Unlike the 7-cell tile, this configuration has an even number of cells and edges. Larger or even-numbered tilings may accommodate different types of multi-cell patterns. In the simulation shown in Fig.~\ref{fig:fig2}(e), using the new 16-cell tile, we took the same parameters as in (d). In this case, the oscillatory state in (d) did not emerge; instead, the system tended to relax to one of a variety of steady states. One such state is the alternating 2-cell polarity pattern shown in (e). Clearly, a pattern of the form in (e) is only possible on an even-numbered tile, suggesting that the oscillations observed in Fig.~\ref{fig:fig2}(d) may result from geometric frustration inherent in the 7-cell tiling.  As our system is complex and highly non-linear, we also see the appearance of other ``disordered'' stable steady states, by which we mean states that do not show a clear coordinated pattern but are nonetheless stable.  An example of one of these states is shown in Fig.~\ref{fig:fig2}(f) and Movie S7, where the polarity of the gels shows abrupt changes between contiguous cells.

We now turn to a systematic  investigation of how polarized and six-fold symmetric states depend on model parameters. Through a numerical continuation procedure, we trace, as a function of the gel nucleation rate $\nu$, the stable steady states that are i) 1-fold symmetric (``polar'' states) and ii) 6-fold symmetric (``hexatic'' states). To characterize these, we compute polar and hexatic order parameters, $\vb p_j$ and $\vb H_j$, in each cell $j$ as:
\eq{
\vb p_j=\frac{1}{M_j}\int_\Omega d^2\vb x\, e^{i\theta(\vb x)}\rho_j(\vb x) \, , \quad \vb H_j=\frac{1}{M_j}\int_\Omega d^2\vb x\, e^{i6\theta(\vb x)}\rho_j(\vb x)\, ;\quad M_j=\int_\Omega d^2\vb x\, \rho_j(\vb x)\, \label{eq:pH},
} 
with $\theta(\vb x)$ representing the angle that $\vb x$ makes with the origin, defined to be the center of the cell, and $\Omega$ representing the 2D domain of the cell. The magnitude of these complex numbers is bounded by 1 and indicates the extent to which the gel density is organized in a polar or hexatic manner, while their phase encodes the associated directionality. We note that for a flat homogeneous state ($\rho=const$) one has $\vb p_j=\vb 0$ but $|\vb H_j| =\bar H \equiv 11/2-\pi\sqrt{3}\approx0.0586$, giving a finite baseline associated with the hexagonal geometry.  
A polar state has $|\vb p_j|>0$ whereas a hexatic state has $\vb p_j=\vb 0$ with $|\vb H_j|>\bar H$. 

\begin{figure}[tbp]
   \centering
   \includegraphics[width=0.9\textwidth]{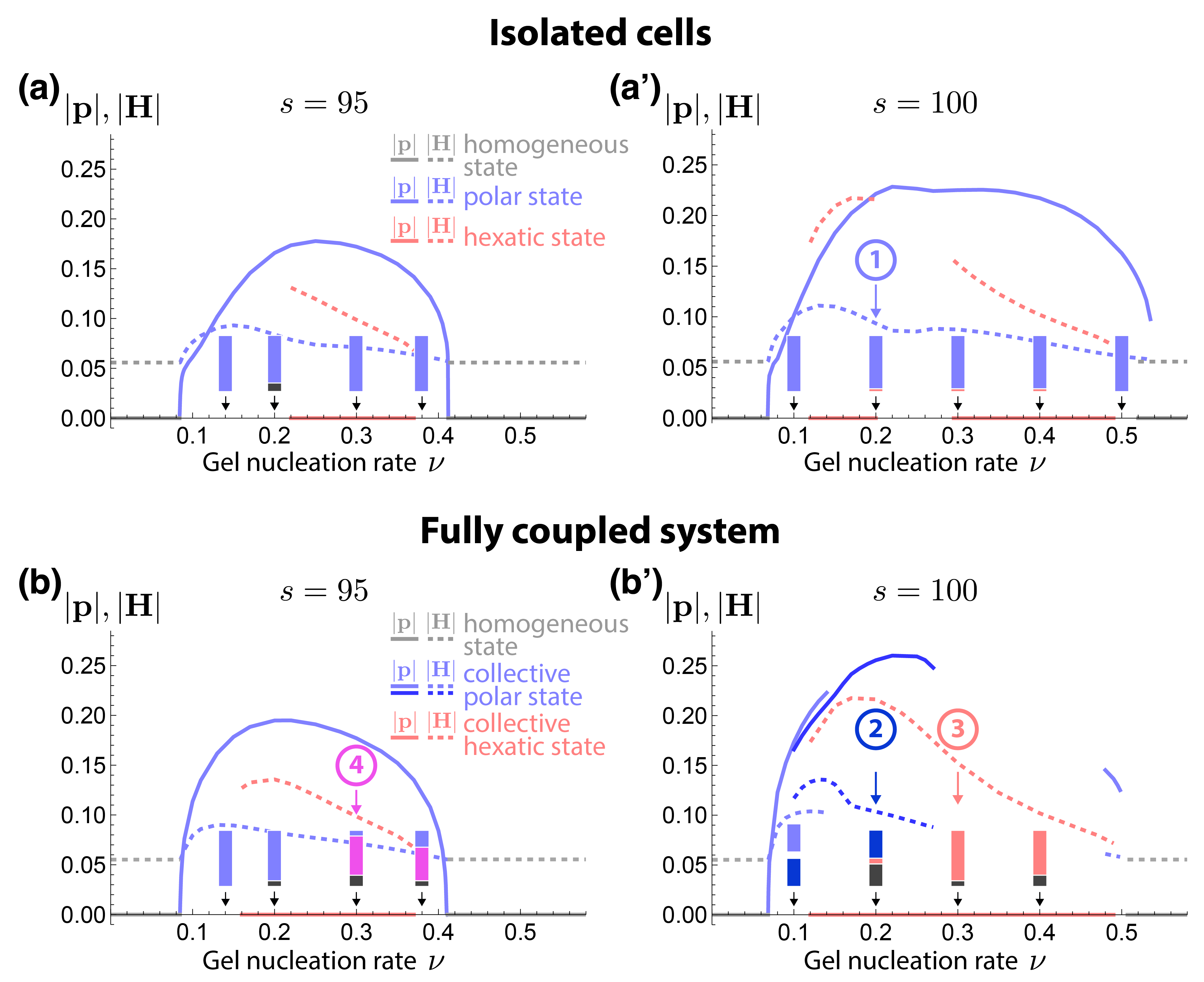} 
   \caption{(a-a') Bifurcation diagram of the homogeneous state (gray), the polar state (blue), and the hexatic state (pink) traced for isolated cells as a function of $\nu$ with other parameters fixed. Column charts indicate the frequency (out of 20 isolated cells) of each stable state to arise in simulation from random initial conditions; polar state (blue bar), hexatic state (pink bar), disordered state (black bar).  (b-b') Bifurcation diagram of the homogeneous state (gray), the collective polar state (blue), and the collective hexatic state (pink) traced for a heptahex system of cells coupled by linkers.  Diagram is a function of $\nu$ with other parameters fixed. Column charts indicate the frequency (out of 10 fully coupled systems) of each stable state to arise in simulation from random initial conditions; collective polar states (blue bar, dark blue bar), collective hexatic states (pink bar), oscillating states (magenta bar), disordered state (black bar). Gel parameters (a-b') $s_1=30$, $\gamma=0.001$, $L_s=8$. Linker parameters (b,b') $a=80$, $\lambda_0=10$, $D_\text{e}=0.1$, $\mu_0=1$, $\ell_\text{M}=0.6$, $\ell_\text{N}=0.1$. }
   \label{fig:fig3}
\end{figure}

For isolated cells, we consider $|\vb p|$ and $|\vb H|$ of single cells. For the fully coupled system, we define $|\vb p|=|\left<\vb p_j\right>|$ and $|\vb H|=|\left<\vb H_j\right>|$, where $\left<\vdot\right>$ denotes the average over all cells in the tile. In collective states where all cells end up in the same configuration, $|\vb p|$ and $|\vb H|$ will be same as those of each individual in the tile.  Figure \ref{fig:fig3} shows $|\vb p|$ and $|\vb H|$ for polar and hexatic stable steady states as the actin nucleation rate $\nu$ is varied. These bifurcation diagrams are found through numerical continuation in $\nu$.  The column charts in Fig.~\ref{fig:fig3} display the fraction of simulations, out of ten runs started from random initial configurations, that settle into different types of steady states (here polar, hexatic, oscillating, and disordered).  The colored and encircled numbers correspond to the states shown in Fig.~\ref{fig:fig2}.  

We focus first on the bifurcation diagram.  For the isolated cells in Fig.~\ref{fig:fig3}(a) with $s=95$, the locally stable homogeneous steady state bifurcates into a polar state in an apparent pitchfork bifurcation as $\nu$ exceeds a critical value and the homogeneous state becomes unstable to condensate formation due to contractility. As $\nu$ increases, the polar branch persists until it undergoes another bifurcation leading to the reentrance of the homogeneous state. An apparently isolated branch of the hexatic state emerges for a range of $\nu$ that overlaps with the polar state, illustrating that this is a multistable system.  For a larger value of the contractility strength $s=100$ in Fig.~\ref{fig:fig3}(a'), an apparent pitchfork bifurcation takes the system from a homogeneous steady state into a polar state at small $\nu$, similar to (a).  But in contrast to (a), the polar state returns to the homogeneous steady state at high $\nu$ through a subcritical bifurcation.  Increasing $s$ to even larger values, shown in Fig.~\ref{S-fig:noecad}, introduces a split in the polar branch.  At the higher contractility strength $s=100$ (a'), the hexatic state splits into two distinct branches. The key qualitative difference between these is whether the gel localizes to the six corners of the cell or forms a continuous ``actin ring'' along the cell boundary.  

A common feature of all the bifurcation diagrams in Fig.~\ref{fig:fig3} is that the system has a flat, homogeneous state that is stable at small and large $\nu$ but is unstable at intermediate values of $\nu$.  This behavior can be understood from the linear stability analysis performed in a periodic 2D system in the Appendix. This analysis shows that the (linkerless) homogeneous state of the gel is linearly unstable for a finite range of $\nu$ whenever $s$ is sufficiently high.  The instability in this range leads to condensate formation in the gel.  This general feature remains true in the hexagonal geometry.

We then investigated the possible variety of states to be found in isolated cells at a few chosen values of $\nu$.  We found that randomly chosen initial configurations most frequently evolve into polar steady states, with hexatic states and states having other symmetries appearing with lower frequency (column charts in Fig.~\ref{fig:fig3}(a,a')).  Though we have not explored this further, the states having other symmetries are presumably parts of other solution branches.

Now we explore the bifurcation diagram when linkers are included at sufficient density.  Figure ~\ref{fig:fig3}(b) shows that for lower contractility ($s = 95$), the bifurcation diagram is similar to that of isolated cells; that is, the homogeneous state bifurcates into a polar state that persists across a finite range of $\nu$ until another bifurcation returns the system to a homogeneous state at high $\nu$. The apparently isolated branch of the stable hexatic state is now found for a broader range of $\nu$ compared to isolated cells.  We emphasize here that the polar and hexatic states in the fully linked system now refer to globally polarized and globally hexatic states, as exemplified by Fig.~\ref{fig:fig2}(b,c).  At higher contractility ($s=100$; Fig.~\ref{fig:fig3}(b')), the polar state bifurcates from the homogeneous state at small $\nu$, similar to the bifurcation in isolated cells (a').  However, the polar state in (b') fractures into two branches, with both branches stable only for low values of $\nu$, unlike in isolated cells.  A separate, very short polar branch appears for high values of $\nu$; this latter polar branch extends through a larger range of $\nu$ at lower densities of linkers and disappears for higher densities of linkers (not shown).  For the hexatic states with $s=100$, the range of stability is extended compared to that in isolated cells; in fact, they extend enough to merge together the two hexatic branches that were present in (a') into a single branch in (b').

Sampling the fully coupled system through simulations from random initial configurations, we find that polar steady states emerge more often when $\nu$ is small (Fig.\ref{fig:fig3}(b,b')). When $\nu$ is large, hexatic states appear at higher contractility ($s=100$), and oscillatory states now appear at the lower contractility ($s=95$). The higher prevalence of hexatic states at large $\nu$ in the linked system (Fig.\ref{fig:fig3}(b')) is somewhat consistent with intuition: greater actin availability allows for wetting of a larger surface area, and linkers tend to promote the spreading of the actin gel along the surface; see Movies S4 and S5.

\begin{figure}[tbp]
   \centering
   \includegraphics[width=1.0\textwidth]{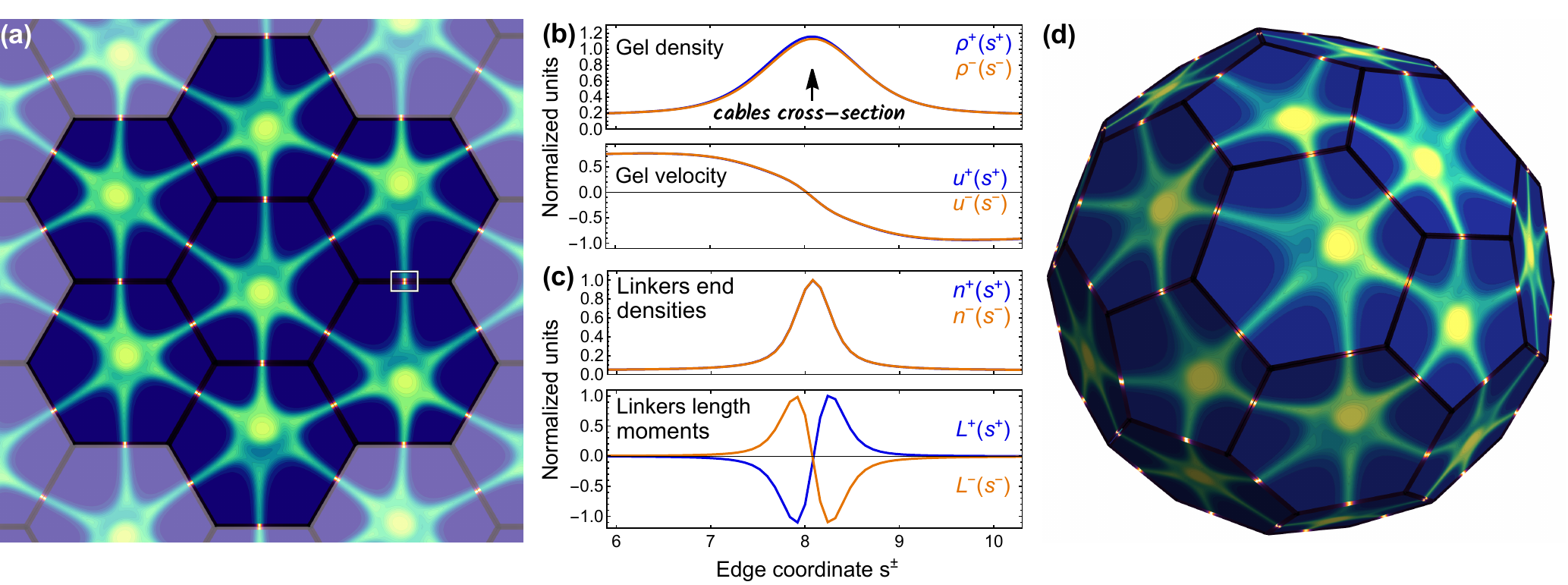} 
   \caption{Supracellular cables. (a) Snapshot from a simulation showing the transient appearance of star-like supracellular cables in the heptahex. Cable junctions at cell–cell interfaces are enriched with linkers (bright red). (b) Cross-sections of gel densities $\rho^\pm(s^\pm)$ and tangential gel velocities $u^\pm(s^\pm)$ along the shared edge of adjacent cells, near the transient cable boxed in (a). (c) Linker end densities $n^\pm(s^\pm)=\int ds^\mp\psi(s^+,s^-)$ and length moments $L^\pm(s^\pm)=\int ds^\mp(s^+-s^-)\psi(s^+,s^-)$ along the same edge as (b). (d) Snapshot from a simulation showing a similar transient cable arrangement emerging in the truncated icosahedron. Gel parameters: $s_1=30$, $\gamma=0.001$, $L_s=8$. Linker parameters: $\lambda_0=10$, $D_\text{e}=0.1$, $\mu_0=1$, $\ell_\text{M}=0.6$, $\ell_\text{N}=0.1$. }
   \label{fig:fig4}
\end{figure}

Beyond the patterns explored so far, a more striking form of organization has been experimentally observed: actomyosin networks in tissues can assemble into dense, cable-like structures spanning multiple cells.  These cables appear to transit continuously across cell boundaries \cite{Yevick2019,Barai2025,Lopez2020}. Since F-actin strands do not actually cross cell membranes, the apparent continuity of cables likely reflects a transmembrane coupling between the actomyosin networks of adjacent cells. We find that our linker-coupled model is capable of reproducing such supracellular cable patterns, at least transiently, when the system size, contractility, and linker density are all sufficiently large; see Fig.~\ref{fig:fig4}. These patterns arise from the mechanical coupling between actomyosin afforded by linkers across the membranes. 

Figure~\ref{fig:fig4}(a) and Movie~S8 show a representative example of supracellular cabling in a 7-cell tile. This simulation shows an evolving star-like pattern of intersecting cables that meet in the cell centers and smoothly cross each shared cell edge. Matching cross-sections of actin density and velocity in the two adjacent cells are shown in Fig.~\ref{fig:fig4}(b). This simulation was started from a nearly homogeneous steady state having a small density perturbation near the center of each cell. The cell-spanning actin cables are maintained by an enriched linker density and specific patterning of linker extensions localized around the peaks in gel density; see Fig.~\ref{fig:fig4}(b,c). Note that in our model, the linker binding rate ($k_\text{on}$) and the actin production rate ($\nu$) are independent of each other's densities, and yet the two components consistently co-localize. Hence, this co-localization arises solely from their mechanical interactions moving their densities together. We find that once a supracellular structure forms spontaneously, the transmembrane cables migrate toward the cell vertices, and the cable eventually disassembles, somewhat resembling experimental observations \cite{Lopez2020}. At long times, the system may either settle into a global steady state or continue to evolve chaotically with actin cables continuously forming and disassembling; see Movie S8.

\FloatBarrier
A fascinating question is how system geometry and topology may impact actomyosin patterning. We have already seen how stable alternating 2-cell patterns cannot exist in the 7-cell tile whereas they can and do exist in the 16-cell tile. This raises the question of what phenomena may emerge in more complex cellular arrangements. Indeed, many biological tissues deviate from perfect hexagonal packing, and in certain developmental contexts, they may close upon themselves in three dimensions. To capture such cases, our framework can be naturally extended to polyhedra with mixed polygonal faces. A representative example is the truncated icosahedron (``soccer ball''), which consists of 32 flat polygonal faces---20 regular hexagons and 12 regular pentagons---connected by 90 shared edges. Since the Euler characteristic of this system is $\chi=2$, a global polarization of all cells as in Fig.~\ref{fig:fig2}(b) is no longer possible, and defects must arise; see Movie S9. This topological obstruction does not prevent the emergence of a star pattern of supracellular cables similar to Fig.~\ref{fig:fig4}(a), as shown in Fig.~\ref{fig:fig4}(d) and Movie S10 (using the same parameters). Simulations on such polyhedra may prove useful to explore supracellular actomyosin patterns in systems such as organoids or cysts that may be engineered \emph{in vitro}.

\section{Discussion}  
Cell–cell communication is central to collective motility, morphogenesis and mechanosensing. The interaction between adhesion complexes and the actin cytoskeleton plays a significant role, but remains poorly understood. A core feature of our model is that actin networks in adjacent cells are mechanically coupled through adhesion complexes that evolve with their own dynamics. This physical perspective on adhesion seems central to understanding how microscopic interactions give rise to large-scale multicellular behavior. We demonstrate that even a minimal mechanical interaction (balancing drag with spring-like restoring forces) is sufficient to generate a rich spectrum of collective behaviors, including steady-state configurations like global polarization and dynamic phenomena such as oscillations and supracellular cables.

Importantly, our theoretical framework can be adapted to test a range of hypotheses about the intrinsic dynamics of actin and linkers, as well as their potential co-regulation.  For example, one can explore the effects of adhesions behaving as either slip- or catch-bonds. For slip-bonds, $k_\text{off}\propto e^{F_\text{sping}/F^*}$, where  $F_\text{spring}=k|s^+-s^-|$ and $F^*$ is a typical force scale, while for catch-bonds, the dependence of $k_\text{off}$ on linker extension is non-monotonic \cite{Buckley2014,Sens2020}; either model can be incorporated by amending Eq.~(\ref{eq:FP}) and (\ref{eq:onoff}).  One may also consider the potential enhancement of actin nucleation by cadherins, as suggested in some studies \cite{Harris2010,Vasioukhin2001}. A straightforward way to incorporate this into our framework is to introduce a $\psi$-dependent flux term in the boundary condition for $\rho$ (altering Eq.~(\ref{eq:rhobc})).  Additionally, one may explore a more complex form for the force between linker ends and actomyosin gels in Eq.~(\ref{eq:eom}).  In fact, such a simple description as Eq.~(\ref{eq:eom}) invites experiment, as the density-dependent drag force and linear spring force are merely reasonable assumptions. Another natural extension of the model would be to take into account the dynamics of non-dimerized adhesion complex components. The interchange between unbound, non-dimerized components that can move throughout the cell and dimerized complexes that are limited to moving along single cell edges might lead to a better understanding of what happens at tri-cellular junctions, where experimentally, adhesion-related proteins have been shown to be enhanced \cite{Vanderleest2018,Yu2020}.   

Finally, this model may be the basis of a free-boundary model that includes the movement and deformation of the cell membrane, enabling exploration of how actomyosin–adhesion coupling drives cell shape changes and collective motility in confluent tissue.  We have shown that supracellular cables emerge from our simulations but that they are short-lived; perhaps the addition of drag force against a dynamic membrane could enable these cables to last longer; in addition, we might explore how cabling often arise in response to cell deformations \cite{Lopez2020}.  The internal active gel dynamics could be enriched by considering a material with polar or nematic order. In living cells, actin filaments in cables and rings are not only denser but also aligned \cite{Barai2025}, and this alignment may induce additional coordination between cells that can be investigated through modeling.  Importantly, as we build an understanding of cell-cell interfaces populated by dynamical assemblies, we can adapt this framework to describe mechanical interactions mediated by transmembrane linkers of any sort, including integrins, in which one linker end experiences forces against cytoskeletal material while another experiences forces against a substrate.  Our system of gels, linkers, and potentially membranes, seems well-suited to model collective tissue behaviors in various environments.


\clearpage
\bibliography{cadherin}{}
\bibliographystyle{unsrt}

\clearpage
\appendix
\renewcommand{\thefigure}{A\arabic{figure}}
\setcounter{figure}{0}  

\section{Linear stability of 2D periodic actomyosin gel}
\begin{figure}[htbp] 
\centering
\includegraphics[width=0.8\textwidth]{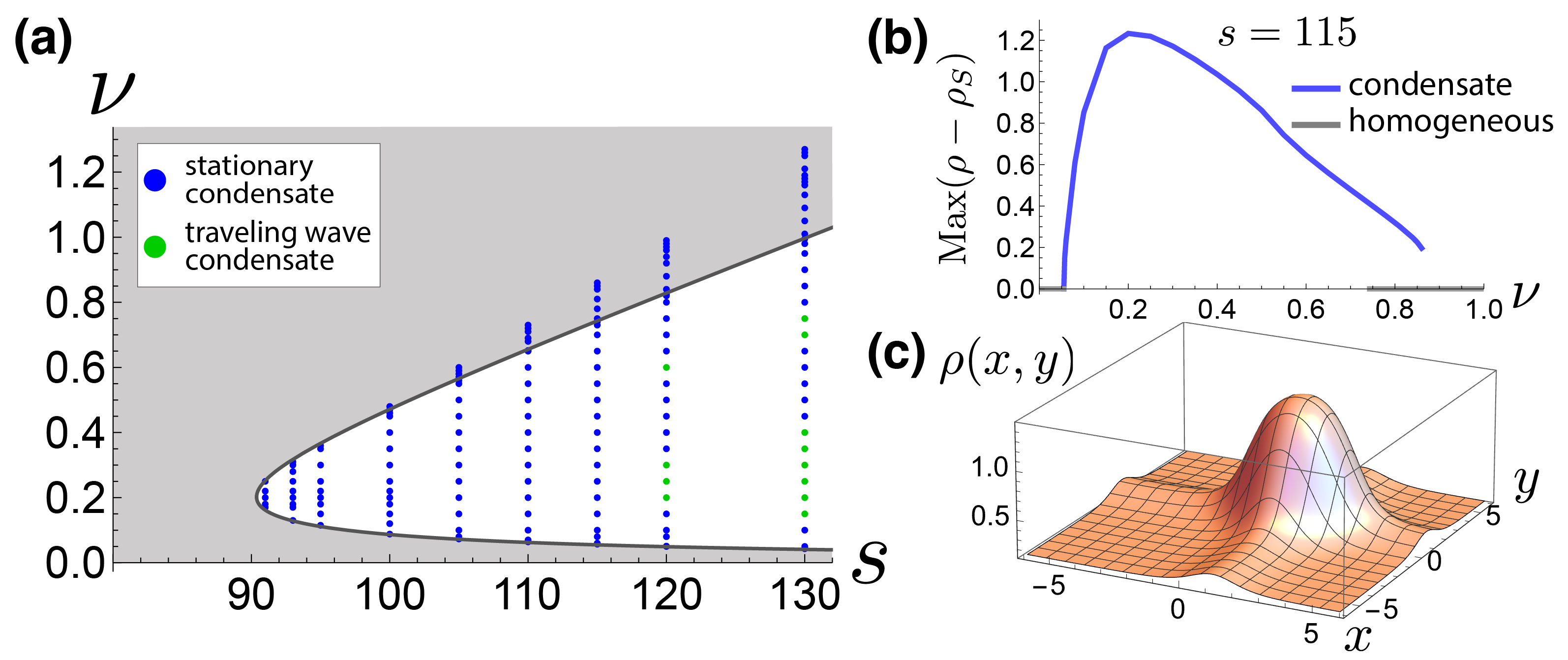} 
   \caption{(a) Stability diagram for 2D actomyosin gel in periodic boundary conditions with varying contractility strength $s$ and gel nucleation $\nu$; other parameters: $s_1=30$, $\gamma=0.001$, $L_s=12$.  The homogeneous state is stable in the gray region.  Condensate states are shown by dots, indicating simulations, with blue dots representing stationary condensate steady states and green dots representing traveling condensate steady states. The overlap of dots with the gray region indicates multistability between the homogeneous state and condensate state. (b) Maximum density $\rho$ above $\rho_S=\nu$ as a function of $\nu$ for $s=115$. Non-zero values of $\text{Max}(\rho-\rho_S)$ indicate condensate states (blue); $\text{Max}(\rho-\rho_S)=0$ indicates homogeneous states. The graph shows multistability between condensate and homogeneous states towards higher $\nu$. (c) Gel density profile $\rho$ for a condensate state with $s=115$, $\nu=0.3$ from the diagram in (a).}
   \label{fig:lin}
\end{figure}

We investigate the linear stability of the homogeneous steady state of the actomyosin gel without linkers in a 2D periodic domain.  Such a system is described by Eqns.~(\ref{eq:rho}-\ref{eq:stress}), reiterated below with Eq.~(\ref{eq:u}) and Eq.~(\ref{eq:stress}) combined into Eq.~(\ref{eq:ustress}):
\aleq{
\pdv{\rho}{t} =& D\laplacian\rho-\div (\vb u\rho) +\nu-b\rho \label{eq:rho_copy} \\
0 =& \div\qty(\eta (\rho+\rho_0)^2 (\grad \vb u+\grad\vb u^T)) - \grad(s_1 (\rho+\rho_0)^2 -s\rho) -\gamma (\rho+\rho_0)\vb u  \label{eq:ustress}
}
The homogeneous steady state of Eqns.~(\ref{eq:rho_copy}) and (\ref{eq:ustress}) are described by states where ${\rho_S=const.}$ and $\vb u_S=\vb 0$, namely:
\eq{
\rho_S = \frac \nu b \quad, \quad \vb u_S=\vb 0 \quad .
}
To perturb the system linearly, we write:
\eq{
\rho \to \rho_S+\delta \rho \quad, \quad \vb u\to \delta \vb u \quad .
}
Then Eq.~(\ref{eq:rho_copy}) and Eq.~(\ref{eq:ustress}) to linear order in $\delta\rho$ and $\delta\vb u$ are:
\aleq{
&\pdv{\delta\rho}{t} = D\laplacian\delta\rho -(\div\delta\vb u) \rho_S -b\delta\rho  \label{eq:rholinear} \\
&0 = \eta(\rho_S +\rho_0)^2 
(\laplacian\delta\vb u + \grad(\div \delta\vb u))-s_1 2(\rho_S+\rho_0)\grad\delta\rho + s\grad\delta\rho - \gamma(\rho_S+\rho_0)\delta\vb u \quad. \label{eq:ulinear}
}
Taking the Fourier transform of Eq.~(\ref{eq:ulinear}), and using $\hat{\delta\rho}$ and $\hat{\delta\vb u}$ to notate the Fourier-transformed fields $\hat{\delta\rho}(\vb k)$ and $\hat{\delta\vb u}(\vb k)$, we have:
\eqlabel{eq:dueq}{
0 =\eta(\rho_S +\rho_0)^2 (-k^2\hat{\delta\vb u}+ i\vb k(i\vb k\cdot \hat{\delta\vb u} ))) 
-s_1 2(\rho_S +\rho_0)(i\vb k)\hat{\delta\rho}+s(i\vb k)\hat{\delta\rho} -\gamma (\rho_S +\rho_0)\hat{\delta\vb u} \quad .
}
Similarly, taking the Fourier Transform of Eq.~(\ref{eq:rholinear}), we have:
\eqlabel{eq:rhofourier}{
\pdv{\hat{\delta\rho}}{t} = -k^2D\hat{\delta\rho} -(i\vb k\cdot\hat{\delta\vb u}) \rho_S -b\hat{\delta\rho} \quad .
}
Here $k\equiv |\vb k|$. The expression $(i\vb k\cdot\hat{\delta\vb u})$ is needed to eliminate $\delta\vb u$ from Eq.~(\ref{eq:rhofourier});  this is obtained by taking the dot product of $\vb k$ with Eq.~(\ref{eq:dueq}) and solving for $(i\vb k\cdot\hat{\delta\vb u})$, which gives:
\eqlabel{eq:usol}{
(i\vb k\cdot \hat{\delta\vb u}) = \frac{k^2(2s_1(\rho_S+\rho_0)-s) \hat{\delta\rho}}{2k^2\eta(\rho_S+\rho_0)^2 + \gamma(\rho_S+\rho_0)} \quad .
}
Substituting in the above into Eq.~(\ref{eq:rhofourier}) and 
defining $F \equiv 2k^2\eta(\rho_S+\rho_0)^2 +\gamma (\rho_S+\rho_0)$ for algebraic simplification, we have:
\eq{
\pdv{\hat{\delta\rho}}{t} =\qty[\frac{-k^2DF - k^2(2 s_1(\rho_S+\rho_0)-s) \rho_S -bF}{F}]\hat{\delta\rho} \quad . \label{eq:drhoeq}
}
Notably, $F$ is positive definite, so the stability of the homogeneous state $\rho=\rho_S$ is determined by the sign of the numerator in the coefficient of $\hat{\delta\rho}$ in Eq.~(\ref{eq:drhoeq}): 
\eqlabel{eq:stab}{
-k^2DF - k^2(2 s_1(\rho_S+\rho_0)-s) \rho_S -bF <0 \quad \implies \text{stability}
}
Since this expression is simply a 2nd order polynomial in $\rho_S$ with a negative squared term equal to $-2 (b\eta k^2 + D\eta k^4 + k^2 s_1) \rho_S^2$, obtained by substituting $F$ into Eq.~\eqref{eq:stab}, it is a concave down parabola in $\rho_S$.  This implies that the homogeneous state becomes unstable for any value of $\rho_S=\nu/b$ (or $\rho_S=\nu$ in non-dimensional units) where this concave down parabola is positive.  Hence, given mode $\vb k$ and non-dimensional parameters $s$, $s_1$, and $\gamma$ and varying $\nu$ (or $\rho_S$), we will find either no values or a finite range of values of $\nu$ for which the homogeneous state is unstable. For small $s$, the homogenous state ($\rho_S=\nu$) is stable for all values of $\nu$. When $s$ is increased past a critical value, the gel forms a condensate for some finite range of $\nu$ where the homogeneous state is unstable; see Fig.~\ref{fig:lin}.

\end{document}